\RequirePackage{lineno} 
\documentclass{nature}
\usepackage{xcolor}

\usepackage{caption}
\usepackage{multirow}
\usepackage{amsmath}
\usepackage{bm}
\usepackage{braket}
\usepackage{amsmath}

\title{Self-energy dynamics and  mode-specific  phonon threshold effect in a Kekul\'e-ordered graphene}

\author{Hongyun Zhang$^1$$^\dagger$, Changhua Bao$^1$$^\dagger$, Michael Sch\"uler$^2$, Shaohua Zhou$^1$, Qian Li$^1$, Laipeng Luo$^1$, Wei Yao$^1$, Zhong Wang$^3$, Thomas P. Devereaux$^{2,4}$ \& Shuyun Zhou$^{1,5,*}$}

\usepackage{graphicx}
\makeatletter
\let\saved@includegraphics\includegraphics
\AtBeginDocument{\let\includegraphics\saved@includegraphics}

\makeatother

\begin{document}
\maketitle

\begin{affiliations}
 \item State Key Laboratory of Low Dimensional Quantum Physics and Department of Physics, Tsinghua University, Beijing 100084, P. R. China
 \item Stanford Institute for Materials and Energy Sciences (SIMES), SLAC National Accelerator Laboratory, Menlo Park, California 94025, USA
 \item Institute for Advanced Study, Tsinghua University, Beijing, 100084, P. R. China
 \item Department of Materials Science and Engineering, Stanford University, Stanford, CA 94035, USA
 \item Frontier Science Center for Quantum Information, Beijing 100084, P. R. China \\
 $\dagger$ These authors contributed equally to this work.
 
 * Correspondence should be sent to syzhou@mail.tsinghua.edu.cn.
\end{affiliations}


\begin{abstract}
	
{\bf Electron-phonon interaction and related self-energy are fundamental to both the equilibrium properties and non-equilibrium relaxation dynamics of solids. Although electron-phonon interaction has been suggested by various time-resolved measurements to be important for the relaxation dynamics of graphene, the lack of energy- and momentum-resolved self-energy dynamics prohibits direct identification of the role of specific phonon modes in the relaxation dynamics. Here by performing time- and angle-resolved photoemission spectroscopy measurements on a Kekul\'e-ordered graphene with folded Dirac cones at the $\Gamma$ point, we have succeeded in resolving the self-energy effect induced by coupling of electrons to two phonons at $\Omega_1$ = 177 meV and $\Omega_2$ = 54 meV and revealing its dynamical change in the time domain.  Moreover, these strongly coupled phonons define energy thresholds, which separate the hierarchical relaxation dynamics from ultrafast, fast to slow, thereby providing direct experimental evidence for the dominant role of mode-specific phonons in the relaxation dynamics. 	
}

\end{abstract}

\newpage

Electron-phonon interaction is ubiquitous in solids and fundamental to the transport properties  \cite{ShamLJ,AshcroftMermin} as well as the non-equilibrium relaxation dynamics. Electron-phonon interaction determines the electrical resistivity of metals \cite{ShamLJ,AshcroftMermin}, affects the electron mobility of semiconductors, and drives phase transitions such as charge density wave \cite{GrunerCDW}, superconductivity \cite{BCS1957,McMillan}. 
Electron-phonon interaction also plays a critical role in the non-equilibrium relaxation dynamics, as has been revealed by various time-resolved optical measurements, where the relaxation rate of electrons is determined by the electron-phonon coupling strength averaged over all phonons  \cite{MourouPRL87,SchoenleinPRL87,AllenPRL87,DresselhausPRL90}.  In order to further identify whether the relaxation dynamics is dominantly determined by specific phonon modes that are strongly coupled with electrons or contributed by all phonons, it is important to experimentally resolve the self-energy $\Sigma$  in the time domain and reveal its dynamic evolution. 
Angle-resolved photoemission spectroscopy (ARPES) is a powerful tool for extracting the real and imaginary parts of the self-energy Re$\Sigma$ and $|$Im$\Sigma|$, which show up  in the ARPES data as a renormalization of the electronic dispersion \cite{AshcroftMermin} and an increase of the scattering rate near the phonon energy. By combining ARPES with ultrafast pump-probe, time-resolved ARPES (TrARPES) provides unique opportunities for revealing the self-energy effect in the time domain with mode-specific information and establishing a direct connection between the strongly coupled phonons and the relaxation dynamics.

\begin{figure*}[]
	\centering
	\includegraphics[width=16.5 cm]{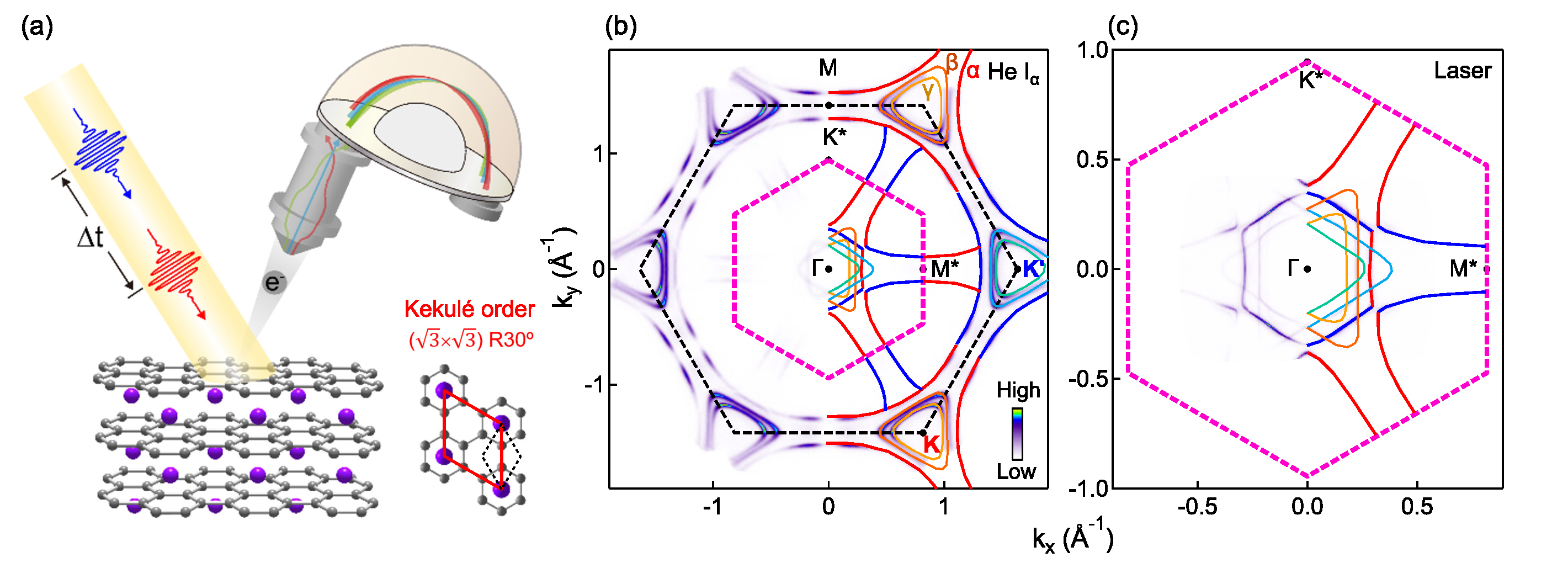}
	\caption{\textbf{A schematic for TrARPES and Fermi surface map of Kekul\'e-ordered graphene.} (a) A schematic of TrARPES on Li-intercalated trilayer graphene with a superlattice period of ($\sqrt3\times\sqrt3$)R30$^\circ$. The unit cells for graphene and the Kekul\'e order are labelled by black and red parallelograms. (b) Fermi surface map of the Li-intercalated trilayer graphene measured with Helium lamp source at 21.2 eV. Red and blue curves indicate the largest pockets ($\alpha$) at K and K$^\prime$. Black and pink broken hexagons are graphene and superlattice BZs respectively. Colored curves around $\Gamma$ indicate folded pockets from K and K$^\prime$ points. (c) Zoomed-in Fermi surface map using 6.2 eV laser source. The largest pockets from the K and K$^\prime$ are highlighted by red and blue colors. The pink broken hexagon is the superlattice BZ with high symmetry points K$^*$ and M$^*$ labelled.}	\label{Fig1}
\end{figure*}

Graphene with low-energy excitations resembling relativistic Dirac fermions \cite{Geim2009Science,Neto2009RMP} and strong electron-phonon coupling indicated by Kohn anomaly \cite{KohnAnomaly}  is a model system for investigating the electron-phonon interaction in both the equilibrium and non-equilibrium states.  
Electron-phonon coupling induced self-energy effect has been resolved in ARPES measurements  of graphene and graphite \cite{ZhouPRB,RotenbergPRL,GruneisNC,GiustinoCaC6phonon,VallaCaC6PRL,ShenCaC6NC,DamascelliPNAS} and suggested to be important for the carrier relaxation from time-resolved optical measurements \cite{WolfGraphitePRL2005,HelmPRL2011,ElsaesserTrOpt2011}. While
TrARPES measurements have been performed to reveal the relaxation dynamics of photo-excited carriers \cite{Cavalleri_2013,Hofmann_2013,Hofmann_2014Nano,Cavalleri_2015,Cavalleri_2014,Hofmann_2014PRL,CavalleriPRL2015Bilayer,Gierz_2017PRB,GierzPRB2017,MatsudaPRB17,PhysRevB.101.035128,BauerPRLG2018,Damascelli2019Sci}, so far, the electron-phonon coupling induced self-energy effect and the associated self-energy dynamics have not been resolved in any TrARPES measurement of graphene or graphite yet.  Experimentally, this is limited by the reduced efficiency and resolution of the high harmonic generation (HHG) light source \cite{GedikNatureComm,Damascelli2019Sci,BauerNature2011}, which is required for generating a sufficiently high photon energy to probe the Dirac cone at the K point with a large momentum value of 1.7 \AA$^{-1}$.  

Here by taking an experimental strategy of folding the Dirac cones from K to $\Gamma$ by inducing a $(\sqrt{3}\times\sqrt{3})R30^\circ$ (Fig.~1(a))  Kekul\'e order \cite{Chamon_2000,Cheianov_2009} through Li intercalation \cite{JohanssonPRB2010,Takahashi2011,ChanghuaPRL2021}, we are able to probe the dynamics of Dirac cones using 6.2 eV photon energy with greatly improved momentum resolution. This leads to successful identification of coupling of electrons to phonons at $\Omega_1$ = 177 meV and $\Omega_2$ = 54 meV in both Re$\Sigma$ and $|$Im$\Sigma|$ in the time domain. Moreover, these two strongly coupled phonons dominate the relaxation dynamics of electrons by setting energy thresholds for the hierarchical relaxation dynamics from ultrafast, fast to slow. Our work reveals the dynamical modification of the electron-phonon coupling induced self-energy effect in the time domain and highlights the dominant role of mode-specific electron-phonon interaction in the non-equilibrium dynamics.

The Kekul\'e-ordered graphene sample is obtained by intercalating Li \cite{JohanssonPRB2010,Takahashi2011,ChanghuaPRL2021} into a bilayer graphene sample grown on SiC substrate. The intercalation releases the bonding between the buffer layer and SiC substrate \cite{OttavianoPRB}, resulting in three weakly interacting graphene layers in AA stacking as schematically shown Fig.~1(a). 
The electronic structure of a Kekul\'e-ordered bilayer graphene has been investigated recently where experimental evidence of chiral symmetry breaking has been provided \cite{ChanghuaPRL2021}. In this work,  we focus on the electronic dynamics of the folded Dirac cones at the $\Gamma$ point by TrARPES measurements using 6.2 eV laser source with a higher repetition rate, which leads to a much higher experimental efficiency and at least three times improvement in the momentum resolution  (see more information about resolution in methods and Supplementary Fig.~S1) compared to previous TrARPES measurements with HHG light source.  Such improvement is critical for  successfully resolving the self-energy effect in the TrARPES measurements.

Figure 1(b) shows the Fermi surface map measured by a Helium lamp source, which contains three large Fermi pockets (indicated by $\alpha$, $\beta$ and $\gamma$, and colored curves) with different sizes around each Brillouin zone (BZ) corner.  The large pocket size indicates large electron doping induced by the intercalated Li.  Folded Dirac cones by the $(\sqrt{3}\times\sqrt{3})R30^\circ$ Kekul\'e superlattice are observed  at the $\Gamma$ point similar to previous work \cite{ChanghuaPRL2021}, and these folded pockets are better resolved in the zoom-in Fermi surface map measured by using  6.2 eV laser source (Fig.~1(c)).  We note that Li-intercalated monolayer graphene does not show folded Dirac cones at the $\Gamma$ point \cite{OttavianoPRB}, and the Li-intercalated trilayer graphene sample shows a much stronger intensity for the folded pockets at the $\Gamma$ point than the Li-intercalated bilayer graphene sample \cite{ChanghuaPRL2021}.  Therefore, a Li-intercalated trilayer graphene sample is used in this TrARPES study.  Considering that the Li-intercalated samples are arranged in the AA stacking with weak interlayer coupling, the physics of Li-intercalated trilayer and bilayer graphene is expected to be similar.

\begin{figure*}[]
	\centering
	\includegraphics[width=15cm]{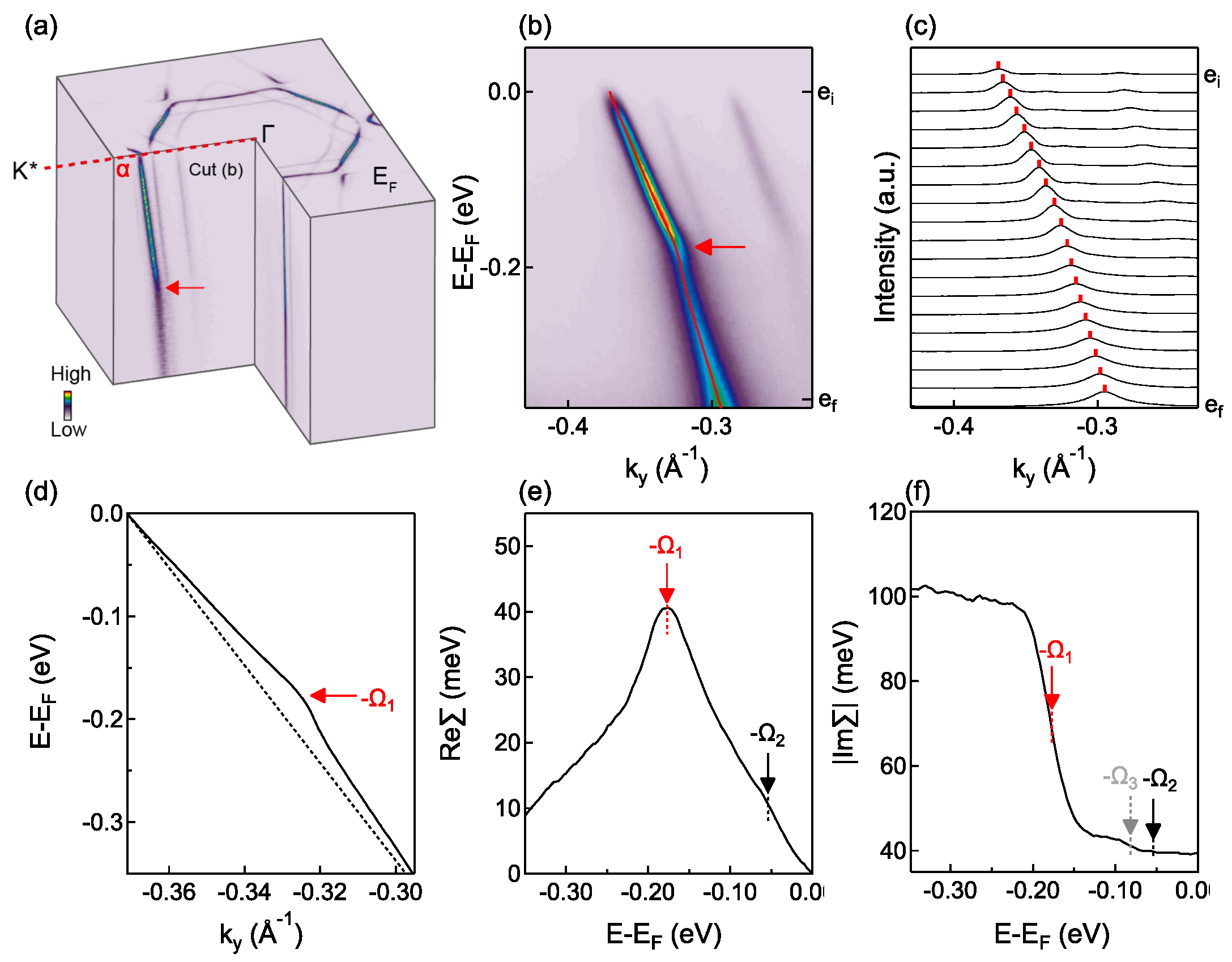}
	\caption{\textbf{Electron-phonon coupling induced self-energy effect.} (a) Three-dimensional band structure of the Kekul\'e-ordered graphene measured by using 6.2 eV laser source. Red arrow indicates the kink, and red broken line indicates the $\Gamma$-K$^*$ ($\Gamma$-M) direction. (b) Dispersion image measured along direction indicated by red broken line in (a) before pump. The dispersing band with the strongest intensity is from the largest FS pocket $\alpha$. (c) MDCs at energy labeled by $e_i$ to $e_f$ in (b), and the red marks indicate the peak positions from Lorentzian fitting. (d) Extracted dispersion from fitting of MDCs in (c). Black dotted line indicates the bare band dispersion used for extracting Re$\Sigma$ in (e). (e,f) Extracted Re$\Sigma$ and $|$Im$\Sigma|$ reveal the coupling of electrons with phonons at energies $\Omega_1$ = 177 and $\Omega_2$ = 54 meV, and $\Omega_3$ = 82 meV (gray arrow).}
	
\end{figure*}

The high-resolution ARPES data allow to resolve the electron-phonon coupling induced self-energy effects in the folded Dirac cones at the $\Gamma$ point. Figure~2(a) shows the three-dimensional band structure measured by the laser source with sharp dispersions, and a kink (indicated by the red arrow) is observed in the dispersion along the $\Gamma$-K$^*$ ($\Gamma$-M) direction (labeled by red broken line in Fig.~2(a)) for the $\alpha$ pocket. The kink is more clearly resolved in the dispersion image in Fig.~2(b). By fitting the momentum distribution curves (MDCs) in Fig.~2(c), we extract the dispersion (black curve in Fig.~2(d)) and the peak width, which can be converted into Re$\Sigma$ and $|$Im$\Sigma|$ using standard ARPES analysis \cite{ZhouPRB}.  Re$\Sigma$ is extracted by assuming a linear bare band dispersion (dotted line in Fig.~2(d)).  A peak at -$\Omega_1$ (red arrow in Fig.~2(e)) and a shoulder at -$\Omega_2$ (black arrow) is observed, which is accompanied by an increase of the scattering rate in $|$Im$\Sigma|$ (Fig.~2(f)), and possible coupling to an additional phonon at $\Omega_3$ (gray arrow)) is also observed. Further fitting of the Eliashberg function gives the phonon energy of $\Omega_1$ = 177 $\pm$ 1 meV  and $\Omega_2$ = 54 $\pm$ 4 meV (see detailed analysis in Supplementary Fig.~S2). We note that electron-phonon coupling has been reported and suggested to be important for superconducting CaC$_6$ \cite{VallaCaC6PRL,ShenCaC6NC} or Li decorated graphene samples \cite{MauriNatPhys2012,DamascelliPNAS}. Here, the high data quality by laser source allows to resolve fine structures in the self-energy, indicating coupling of electrons to multiple phonons. Such mode-specific electron-phonon interaction lays an important foundation for further investigating the role of these phonons in the relaxation dynamics.

\begin{figure*}[]
	\centering
	\includegraphics[width=16.8 cm]{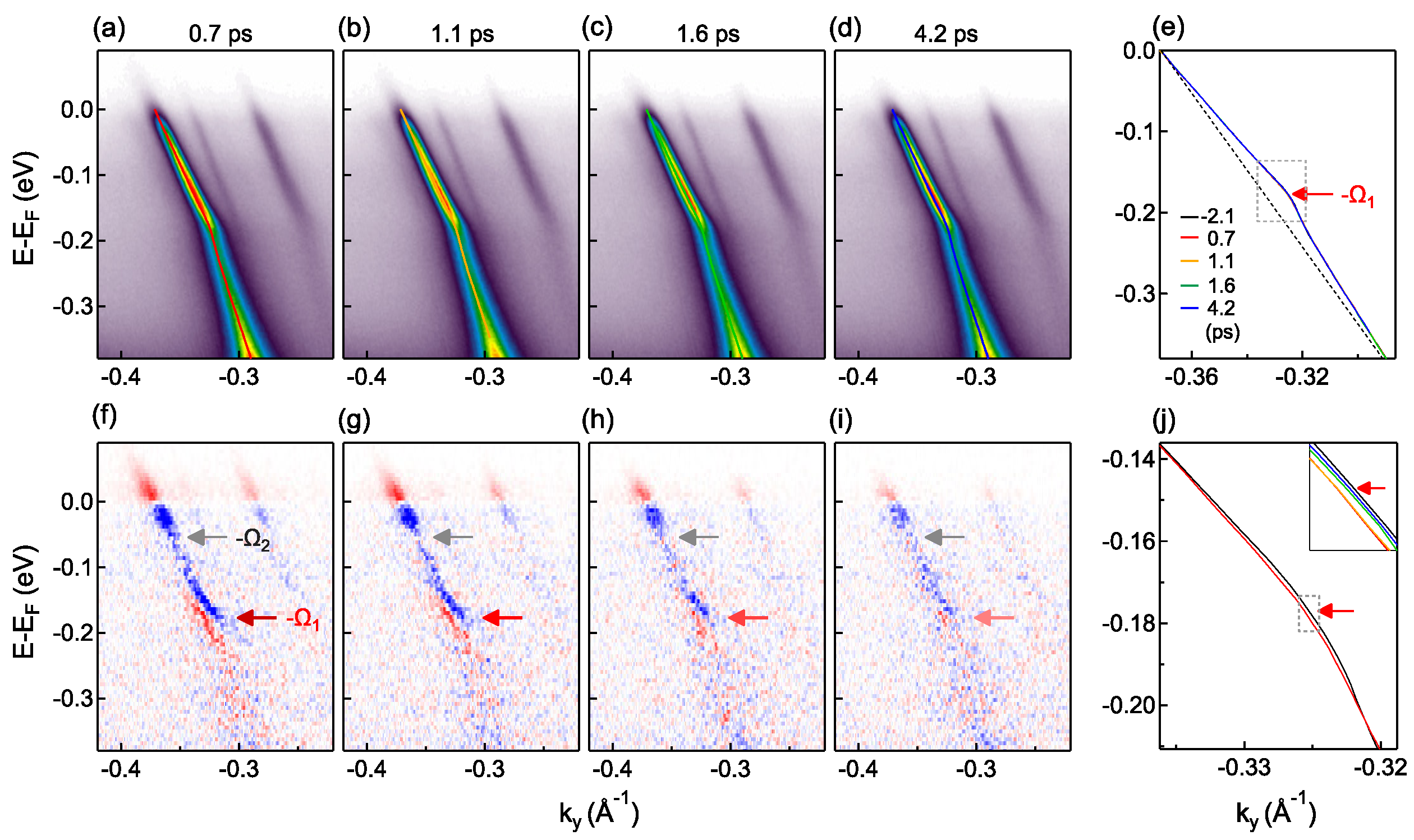}
	\caption{\textbf{TrARPES dispersion images measured at different delay times after pumping at a pump fluence of  215 $\mu$J/cm$^2$.} (a-d) Evolution of the dispersion images measured at different delay times. (e) Extracted dispersion at different delay times. (f-i) Differential images obtained by subtracting dispersion image measured at -2.1 ps from (a-d). Red and blue colors represent increase and decrease of the intensity, respectively. Red and gray arrows indicate the threshold effect at energies of -$\Omega_{1}$ and -$\Omega_{2}$. (j) Zoom-in of the extracted dispersions (indicated by gray dotted box in (e)) to show the comparison of dispersions at 0.7 ps (red curve) and -2.1 ps (black curve). The inset shows the zoom-in dispersion (over the gray dotted box in (j)) at different delay times to show the dynamic evolution of the kink. Red arrows indicate energy of -$\Omega_{1}$.}
\end{figure*}

The electron dynamics is revealed by comparing dispersion images measured at different delay times with a pump photon energy of 1.55 eV at a pump fluence of 215 $\mu$J/cm$^2$. Figure~3(a-d) shows dispersion images measured at 0.7, 1.1, 1.6 and 4.2 ps after pump excitation respectively, and the extracted dispersions at different delay times are plotted in Fig.~3(e). The spectral weight redistribution with an increase of intensity above E$_F$ (red color) indicating photo-excited electrons above E$_F$ and a suppression of intensity below E$_F$ (blue color) indicating photo-excited holes below E$_F$ is clearly resolved in the differential images shown in Fig.~3(f-i) after subtracting the dispersion image measured at -2.1 ps. In contrast to previous TrARPES studies where the TrARPES signal is widely spread out in a large energy range of approximately 1 eV \cite{Cavalleri_2013,Hofmann_2013,Hofmann_2014Nano,Cavalleri_2015,Cavalleri_2014,Hofmann_2014PRL,CavalleriPRL2015Bilayer,Gierz_2017PRB,MatsudaPRB17}, our TrARPES signal is mostly confined in a much smaller energy range within 177 meV (indicated by red arrows) with a much stronger TrARPES signal observed within 54 meV (gray arrows), indicating energy threshold effect in the TrARPES signal and the relaxation dynamics.  Such energy threshold effect with TrARPES signal confined within 177 meV is ubiquitous across the entire BZ (see more data in a larger momentum space in Supplementary Fig.~S3).  In addition, the differential images in Fig.~3(f-i) show an unusual fine feature indicated by red arrows around -$\Omega_1$ with an increase of intensity (red color) at the negative side of the peak and a decrease of intensity at the positive side (blue color), indicating a modification of dispersions measured at different delay times. The zoom-in dispersions in Fig.~3(j) further reveal the dynamical modification of the dispersion near the kink energy at different delay times (see Supplementary Fig.~S4 for a detailed analysis of the self-energy at -2.1 ps and 0.7 ps). At later delay time, the dispersion almost recovers (see comparison of blue curve at -4.2 ps and black curve at -2.1 ps in the inset of Fig.~3(j)).

\begin{figure*}[]
	\centering
	\includegraphics[width=15 cm]{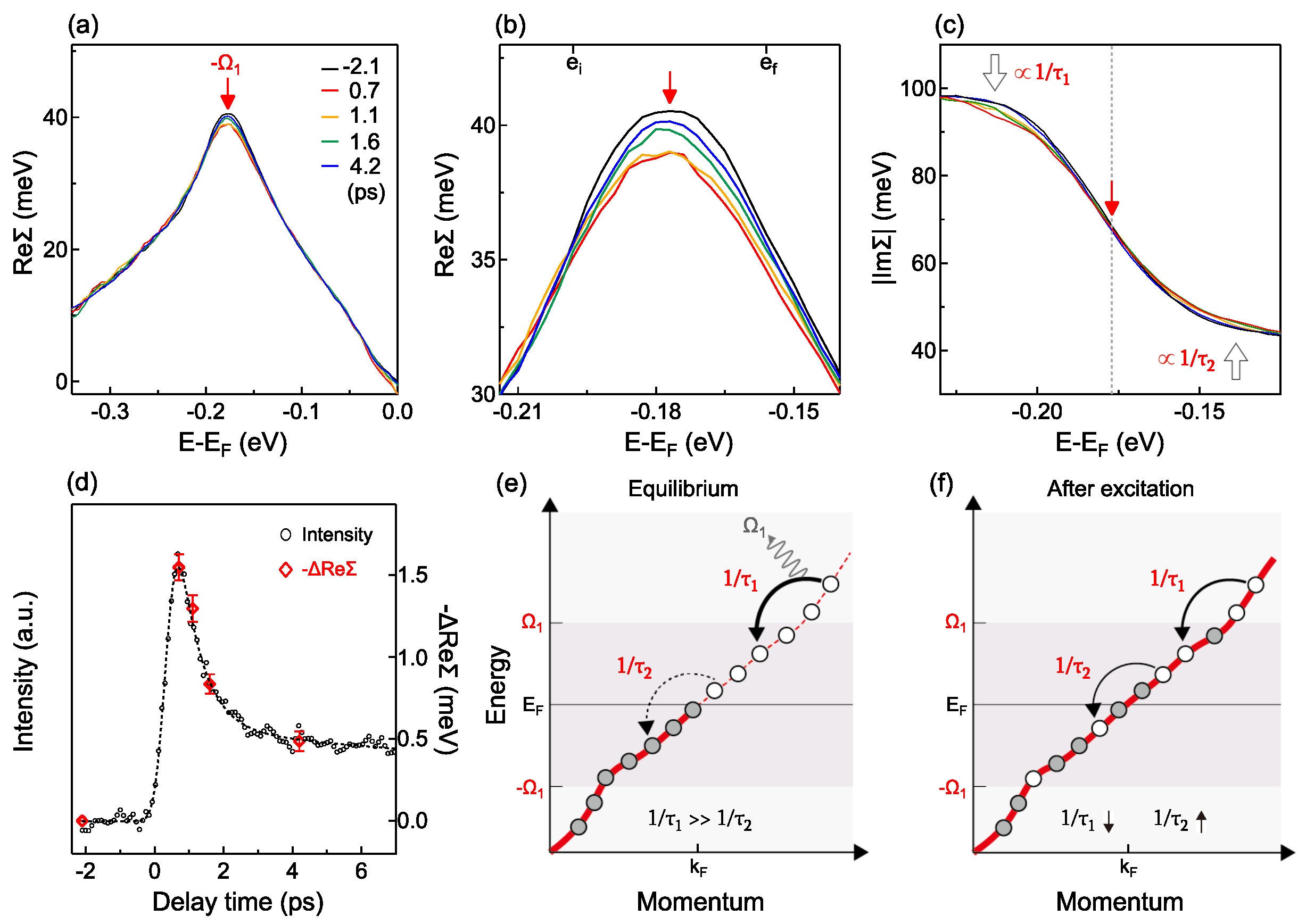}
	\caption{\textbf{Self-energy dynamics and carrier relaxation dynamics.}  (a) Extracted Re$\Sigma$ at different delay times. (b) Zoom-in of the Re$\Sigma$ to show the dynamical change of Re$\Sigma$ around -$\Omega_1$ (red arrow). (c) Extracted $|$Im$\Sigma|$ at different delay times to show the renormalization of the scattering rate around -$\Omega_1$ after pump excitation as indicated by black open arrows. (d) A comparison of pump induced decrease of -Re$\Sigma$ as a function of delay time (red symbols, obtained by averaging the -Re$\Sigma$ over energy range (e$_i$ to e$_f$) indicated by black short marks in (b)), and the pump induced population (black symbols and dotted curve) obtained by integrating from 0 to 50 meV above the Fermi energy. (e,f) Schematics of the electron redistribution after pump excitation and the related renormalization of the scattering rate 1/$\tau$ for holes (or electrons) inside and outside the phonon-window, which gives the dynamical change of the self-energy around the phonon energy. Gray and white balls represent electrons and holes respectively.}
\end{figure*}

To further reveal the underlying physics behind the dynamical change of the dispersion in the time domain, we show in Fig.~4 an analysis of Re$\Sigma$ and $|$Im$\Sigma|$ at different delay times. A decrease of the peak is observed  in Re$\Sigma$ in Fig.~4(a) and the zoom-in Re$\Sigma$ near the kink energy in Fig.~4(b), which gradually recovers at a later delay time (from orange to blue curves in Fig.~4(b)). A corresponding change is also observed in $|$Im$\Sigma|$ (pointed by black open arrows in Fig.~4(c)), which is related to Re$\Sigma$ by the Kramers-Kronig relationship. To check if the electron-phonon coupling induced self-energy effect is correlated with the pump-induced spectral weight transfer revealed in Fig.~3(f-i), we show in Fig.~4(d) a comparison of the temporal evolution of the pump induced change in the self-energy -$\Delta$Re$\Sigma$ (red symbol) and the electron population above $E_F$ which is obtained by integrating the TrARPES intensity from 0 to 50 meV (black symbols and dotted curve). The same temporal evolution suggests a correlation between the dynamical self-energy and the pump-induced spectral weight redistribution and the corresponding change in the scattering phase space  \cite{Devereaux_PRX2013,Tom_PRB2014}. In the equilibrium state (Fig.~4(e)), the scattering rate for electrons (holes) inside the phonon energy window ($\pm\Omega_1$, 0)  1/$\tau_2$ is much less  than that outside this window 1/$\tau_1$ due to the insufficient energy to emit a phonon at $\Omega_1$, as is indicated by the jump in $|$Im$\Sigma|$ (Fig.~2(f)). 
Upon pump excitation, electrons are populated above the Fermi energy \emph{E}$_F$ and holes below \emph{E}$_F$ (indicated by gray and white balls on the red curve in Fig.~4(f)), therefore, the scattering rate for holes (or electrons) inside the phonon window (1/$\tau_2$) increases due to an increase in the scattering phase space to scatter into, while that outside the phonon window (1/$\tau_1$) decreases. Such change of the scattering rate by the photon-induced spectral weight redistribution leads to a dynamical modification of the $|$Im$\Sigma|$ in Fig.~4(c) and Re$\Sigma$ in Fig.~4(b), implying the significant role of the phonons that are coupled to electrons. 

\begin{figure*}[]
	\centering
	\includegraphics[width=16 cm]{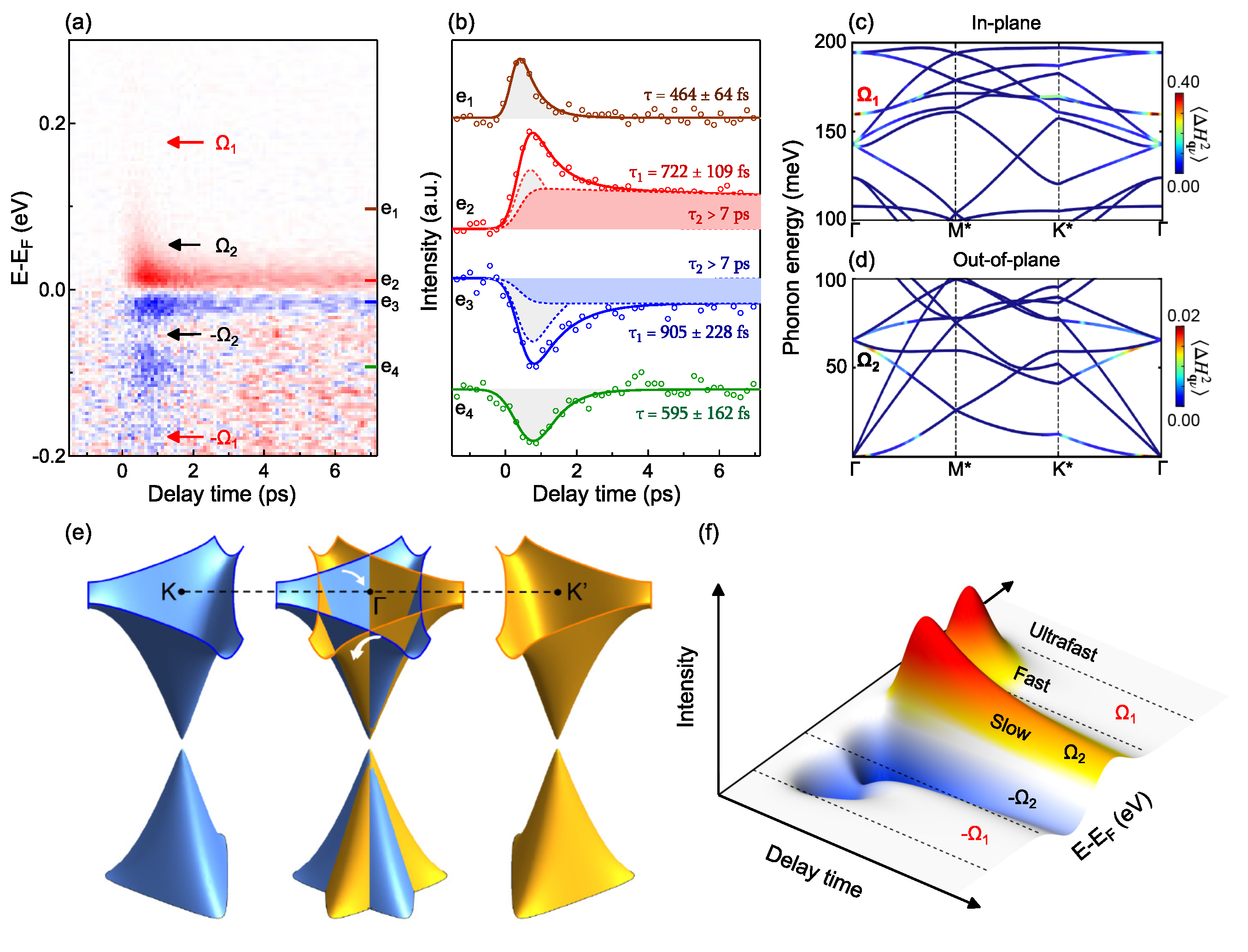}
	\caption{\textbf{Phonon threshold effect and hierarchical relaxation of electrons in different energy windows.} (a) Evolution of momentum-integrated differential intensity with energy and delay time.  Red and blue colors represent increase and decrease of intensity, respectively. (b) Differential intensity as a function of delay time at energies indicated by colored tick marks in (a). Solid curves are fitting results. Gray shadings in e$_1$-e$_4$ indicate the fast relaxation, while red and blue shadings in e$_2$ and e$_3$ represent the slow relaxation component. (c,d) Calculated phonon dispersion and electron-phonon coupling strength of Li-intercalated graphene for in-plane and out-of-plane phonon modes respectively. (e) Schematics of electron-phonon coupling in the Li-intercalated graphene. (f) Schematic drawing for phonon threshold effect with hierarchical relaxation.}
\end{figure*}

The electron-phonon interaction not only modifies the self-energy in the time domain, but also sets energy thresholds as indicated by red and gray arrows in Fig.~3(f-i) and its relation to the relaxation dynamics is further revealed in Fig.~5.  The temporal evolution as a function of energy and delay time (Fig.~5(a)) and the selected curves at different energies (Fig.~5(b)) reveal hierarchical relaxation times in different energy windows defined by the two strongly-coupled phonons, which are summarized below: (1) for energy windows $\pm(\infty, \Omega_1)$ in which photo-excited carriers have sufficient energy to emit phonons both at $\Omega_1$ and $\Omega_2$, the relaxation is ``ultrafast" - faster than 337 fs (see Supplementary  Fig.~S5 for a detailed analysis of energy-dependent relaxation time) and there is negligible TrARPES signal; (2) for energy windows $\pm(\Omega_1, \Omega_2)$ in which photo-excited carriers can emit phonons at $\Omega_2$ but not $\Omega_1$, the relaxation is ``fast", within a few hundred fs (curves at e$_1$ and e$_4$ in Fig.~5(b)); (3) for energy windows $\pm(\Omega_2, 0)$ in which the relaxation requires involvement of acoustic phonons at even lower energy, the relaxation is ``slow" with an additional component persisting beyond 7 ps (highlighted by red and blue shadings for curves e$_2$ and e$_3$ in Fig.~5(b)) which involves relaxation through other mechanism, e.g. acoustic phonons. The observation of distinct relaxation time scales in different energy regimes $\pm(\infty,\Omega_1)$, $\pm(\Omega_1,\Omega_2)$ and $\pm(\Omega_2,0)$ establishes a direct correlation between the hierarchical relaxation times and the two strongly coupled phonons at $\pm(\Omega_1,0)$ and $\pm(\Omega_2,0)$. Therefore, our results show that the coupled phonons $\Omega_1$ and $\Omega_2$ play a dominant role in the relaxation of electrons in graphene.  
Theoretical calculations of the phonon dispersion and electron-phonon coupling strength for the Li-intercalated graphene  (see Supplementary for details of the calculation and Fig.~S6 for more data) have identified that the two phonons $\Omega_1$ and $\Omega_2$ that are coupled to electrons and thereby dominate the relaxation dynamics are the in-plane TO phonon $A_{1g}$ (Fig.~5(c)) and the out-of-plane ZA phonon (Fig.~5(d)).

To summarize, by strategically folding the Dirac cones to $\Gamma$ (Fig.~5(e)), high-resolution TrARPES measurements allow to visualize the coupling of electrons to two strongly coupled phonon modes in the time domain. The electron-phonon interaction not only modifies the electron self-energy, but also sets energy thresholds with hierarchical relaxation dynamics as schematically illustrated in Fig.~5(f). Our work not only provides direct experimental evidence for the dominant role of mode-specific phonons in the relaxation dynamics of a Kekul\'e-ordered graphene, but also provides a new material platform for exploring the engineering of Dirac cones by light-matter interaction.

\begin{methods}
\subsection{\bf Sample preparation}\label{subsec4}

Bilayer graphene was grown by flash annealing the Si face of 6H-SiC(0001) substrates in ultrahigh vacuum. Lithium intercalation was performed by {\it in situ} deposition of Li from an alkali metal dispenser (SAES), with the graphene sample maintained at 320 K \cite{ChanghuaPRL2021}. 
The intercalation process was monitored by low energy electron diffraction (LEED) and ARPES measurements. The intercalation releases the buffer layer \cite{OttavianoPRB} underneath the bilayer graphene, eventually resulting in a Kekul\'e-ordered trilayer graphene with Li atoms inserted between the graphene layers. 

\subsection{\bf TrARPES measurements}\label{subsubsec3}

TrARPES measurements were performed in the home laboratory at Tsinghua University at 80 K in a working vacuum better than $6\times10^{-11}$ Torr. The pump photon energy is 1.55 eV and the pump fluence was set to 215 $\mu$J/cm$^2$. Pulsed laser source at 6.2 eV with a pulse duration of 130 fs and repetition rate of 3.8 MHz is used as the probe source. The overall time resolution was set to 480 fs.   The Fermi edge of the graphene sample measured at 80 K shows an energy width of 33 meV, from which the overall instrumental energy resolution is extracted to be 16 meV after removing the thermal broadening  (see  more details in Supplementary Fig.~S1). Moreover, the reduction of photon energy compared to HHG also leads to major improvement in the momentum resolution. Since the momentum resolution at the Fermi energy \emph{E}$_F$ is $\Delta k \propto \sqrt{h\nu-\phi}$, where \emph{h}$\nu$ and $\phi\approx4.3$ eV are the photon energy and work function respectively, the reduction of photon energy from $h\nu$ $\ge$ 25 eV to 6.2 eV leads to at least three times improvement in $\Delta k$, with an ultimate experimental resolution of $\Delta k$ = 0.001 \AA$^{-1}$.  The greatly improved energy, momentum resolution together with the high data acquisition efficiency is critical for the successful observation of the electron-phonon coupling induced kink in the TrARPES data and the phonon threshold effect.

\end{methods}


\begin{addendum}
\item This work is supported by the National Natural Science Foundation of China (Grant No.~11725418, 11427903),  National Key R $\&$ D Program of China (Grant No.~2016YFA0301004, 2020YFA0308800), Tsinghua University Initiative Scientific Research Program and Tohoku-Tsinghua Collaborative Research Fund, Beijing Advanced Innovation Center for Future Chip (ICFC). M. S. and T. P. D. acknowledge financial support from the U. S. Department of Energy (DOE), Office of Basic Energy Sciences, Division of Materials Sciences and Engineering, under contract No. DE-AC02-76SF00515. M. S. thanks the Alexander von Humboldt Foundation for its support with a Feodor Lynen scholarship.

\item[Author Contributions] Shuyun Z. conceived the research project. H.Z., C.B., Shaohua Z., Q.L., and L.L. performed the TrARPES measurements and analyzed the data. C.B. L.L., and W.Y. grew the graphene samples. M.S. and T.P.D. performed the calculations. Z. W. involved in the discussion. H.Z. and Shuyun Z. wrote the manuscript, and all authors commented on the manuscript.

\item[Author Information]

\item[Competing Interests] The authors declare that they have no competing financial interests.

\item[Correspondence] Correspondence and requests for materials should be addressed to Shuyun Zhou~(email: syzhou@mail.tsinghua.edu.cn).
\end{addendum}


\begin{thebibliography}{10}
	\expandafter\ifx\csname url\endcsname\relax
	\def\url#1{\texttt{#1}}\fi
	\expandafter\ifx\csname urlprefix\endcsname\relax\def\urlprefix{URL }\fi
	\providecommand{\bibinfo}[2]{#2}
	\providecommand{\eprint}[2][]{\url{#2}}
	
	\bibitem{ShamLJ}
	\bibinfo{author}{Sham, L.~J.} \& \bibinfo{author}{Ziman, J.~M.}
	\newblock \bibinfo{title}{The electron-phonon interaction}.
	\newblock \emph{\bibinfo{journal}{{Solid State Phys.}}}
	\textbf{\bibinfo{volume}{15}}, \bibinfo{pages}{221--298}
	(\bibinfo{year}{1963}).
	
	\bibitem{AshcroftMermin}
	\bibinfo{author}{Ashcroft, N.~W.} \& \bibinfo{author}{Mermin, N.}
	\newblock \emph{\bibinfo{title}{Solid state physics.}}
	(\bibinfo{publisher}{Saunders College: Philadelphia}, \bibinfo{year}{1976}).
	
	\bibitem{GrunerCDW}
	\bibinfo{author}{Gr\"uner, G.}
	\newblock \bibinfo{title}{The dynamics of charge-density waves}.
	\newblock \emph{\bibinfo{journal}{Rev. Mod. Phys.}}
	\textbf{\bibinfo{volume}{60}}, \bibinfo{pages}{1129} (\bibinfo{year}{1988}).
	
	\bibitem{BCS1957}
	\bibinfo{author}{Bardeen, J.}, \bibinfo{author}{Cooper, L.~N.} \&
	\bibinfo{author}{Schrieffer, J.~R.}
	\newblock \bibinfo{title}{Theory of superconductivity}.
	\newblock \emph{\bibinfo{journal}{Phys. Rev.}} \textbf{\bibinfo{volume}{108}},
	\bibinfo{pages}{1175} (\bibinfo{year}{1957}).
	
	\bibitem{McMillan}
	\bibinfo{author}{McMillan, W.~L.}
	\newblock \bibinfo{title}{Transition temperature of strong-coupled
		superconductors}.
	\newblock \emph{\bibinfo{journal}{Phys. Rev.}} \textbf{\bibinfo{volume}{167}},
	\bibinfo{pages}{331} (\bibinfo{year}{1968}).
	
	\bibitem{MourouPRL87}
	\bibinfo{author}{Elsayed-Ali, H.~E.}, \bibinfo{author}{Norris, T.~B.},
	\bibinfo{author}{Pessot, M.~A.} \& \bibinfo{author}{Mourou, G.~A.}
	\newblock \bibinfo{title}{Time-resolved observation of electron-phonon
		relaxation in copper}.
	\newblock \emph{\bibinfo{journal}{Phys. Rev. Lett.}}
	\textbf{\bibinfo{volume}{58}}, \bibinfo{pages}{1212} (\bibinfo{year}{1987}).
	
	\bibitem{SchoenleinPRL87}
	\bibinfo{author}{Schoenlein, R.~W.}, \bibinfo{author}{Lin, W.~Z.} \&
	\bibinfo{author}{Fujimoto, J.~G.}
	\newblock \bibinfo{title}{Femtosecond studies of nonequilibrium electronic
		processes in metals}.
	\newblock \emph{\bibinfo{journal}{Phys. Rev. Lett.}}
	\textbf{\bibinfo{volume}{58}}, \bibinfo{pages}{1680} (\bibinfo{year}{1987}).
	
	\bibitem{AllenPRL87}
	\bibinfo{author}{Allen, P.~B.}
	\newblock \bibinfo{title}{Theory of thermal relaxation of electrons in metals}.
	\newblock \emph{\bibinfo{journal}{Phys. Rev. Lett.}}
	\textbf{\bibinfo{volume}{59}}, \bibinfo{pages}{1460} (\bibinfo{year}{1987}).
	
	\bibitem{DresselhausPRL90}
	\bibinfo{author}{Brorson, S.~D.} \emph{et~al.}
	\newblock \bibinfo{title}{Femtosecond room-temperature measurement of the
		electron-phonon coupling constant \ensuremath{\gamma} in metallic
		superconductors}.
	\newblock \emph{\bibinfo{journal}{Phys. Rev. Lett.}}
	\textbf{\bibinfo{volume}{64}}, \bibinfo{pages}{2172} (\bibinfo{year}{1990}).
	
	\bibitem{Geim2009Science}
	\bibinfo{author}{Geim, A.~K.}
	\newblock \bibinfo{title}{Graphene: status and prospects}.
	\newblock \emph{\bibinfo{journal}{Science}} \textbf{\bibinfo{volume}{324}},
	\bibinfo{pages}{1530} (\bibinfo{year}{2009}).
	
	\bibitem{Neto2009RMP}
	\bibinfo{author}{Neto, A. H.~C.}, \bibinfo{author}{Guinea, F.},
	\bibinfo{author}{Peres, N. M.~R.}, \bibinfo{author}{Novoselov, K.~S.} \&
	\bibinfo{author}{Geim, A.~K.}
	\newblock \bibinfo{title}{The electronic properties of graphene}.
	\newblock \emph{\bibinfo{journal}{Rev. Mod. Phys.}}
	\textbf{\bibinfo{volume}{81}}, \bibinfo{pages}{109} (\bibinfo{year}{2009}).
	
	\bibitem{KohnAnomaly}
	\bibinfo{author}{Piscanec, S.}, \bibinfo{author}{Lazzeri, M.},
	\bibinfo{author}{Mauri, F.}, \bibinfo{author}{Ferrari, A.~C.} \&
	\bibinfo{author}{Robertson, J.}
	\newblock \bibinfo{title}{Kohn anomalies and electron-phonon interactions in
		graphite}.
	\newblock \emph{\bibinfo{journal}{Phys. Rev. Lett.}}
	\textbf{\bibinfo{volume}{93}}, \bibinfo{pages}{185503}
	(\bibinfo{year}{2004}).
	
	\bibitem{ZhouPRB}
	\bibinfo{author}{Zhou, S.~Y.}, \bibinfo{author}{Siegel, D.~A.},
	\bibinfo{author}{Fedorov, A.~V.} \& \bibinfo{author}{Lanzara, A.}
	\newblock \bibinfo{title}{Kohn anomaly and interplay of electron-electron and
		electron-phonon interactions in epitaxial graphene}.
	\newblock \emph{\bibinfo{journal}{Phys. Rev. B}} \textbf{\bibinfo{volume}{78}},
	\bibinfo{pages}{193404} (\bibinfo{year}{2008}).
	
	\bibitem{RotenbergPRL}
	\bibinfo{author}{McChesney, J.~L.} \emph{et~al.}
	\newblock \bibinfo{title}{Extended van {H}ove singularity and superconducting
		instability in doped graphene}.
	\newblock \emph{\bibinfo{journal}{Phys. Rev. Lett.}}
	\textbf{\bibinfo{volume}{104}}, \bibinfo{pages}{136803}
	(\bibinfo{year}{2010}).
	
	\bibitem{GruneisNC}
	\bibinfo{author}{Fedorov, A.~V.} \emph{et~al.}
	\newblock \bibinfo{title}{Observation of a universal donor-dependent
		vibrational mode in graphene}.
	\newblock \emph{\bibinfo{journal}{Nat. Commun.}} \textbf{\bibinfo{volume}{5}},
	\bibinfo{pages}{3257} (\bibinfo{year}{2014}).
	
	\bibitem{GiustinoCaC6phonon}
	\bibinfo{author}{Margine, E.~R.}, \bibinfo{author}{Lambert, H.} \&
	\bibinfo{author}{Guistino, F.}
	\newblock \bibinfo{title}{Electron-phonon interaction and pairing mechanism in
		superconducting {C}a-intercalated bilayer graphene}.
	\newblock \emph{\bibinfo{journal}{Sci. Rep.}} \textbf{\bibinfo{volume}{6}},
	\bibinfo{pages}{21414} (\bibinfo{year}{2016}).
	
	\bibitem{VallaCaC6PRL}
	\bibinfo{author}{Valla, T.} \emph{et~al.}
	\newblock \bibinfo{title}{{Anisotropic electron-phonon coupling and dynamical
			nesting on the graphene sheets in superconducting CaC$_6$ using
			angle-resolved photoemission spectroscopy}}.
	\newblock \emph{\bibinfo{journal}{Phys. Rev. Lett.}}
	\textbf{\bibinfo{volume}{102}}, \bibinfo{pages}{107007}
	(\bibinfo{year}{2009}).
	
	\bibitem{ShenCaC6NC}
	\bibinfo{author}{Yang, S.-L.} \emph{et~al.}
	\newblock \bibinfo{title}{Superconducting graphene sheets in {C}a{C}$_6$
		enabled by phonon-mediated interband interactions}.
	\newblock \emph{\bibinfo{journal}{Nat. Commun.}} \textbf{\bibinfo{volume}{5}},
	\bibinfo{pages}{3493} (\bibinfo{year}{2014}).
	
	\bibitem{DamascelliPNAS}
	\bibinfo{author}{Ludbrook, B.~M.} \emph{et~al.}
	\newblock \bibinfo{title}{Evidence for superconductivity in {L}i-decorated
		monolayer graphene}.
	\newblock \emph{\bibinfo{journal}{Proc. Natl. Acad. Sci.}}
	\textbf{\bibinfo{volume}{112}}, \bibinfo{pages}{11795}
	(\bibinfo{year}{2015}).
	
	\bibitem{WolfGraphitePRL2005}
	\bibinfo{author}{Kampfrath, T.}, \bibinfo{author}{Perfetti, L.},
	\bibinfo{author}{Schapper, F.}, \bibinfo{author}{Frischkorn, C.} \&
	\bibinfo{author}{Wolf, M.}
	\newblock \bibinfo{title}{Strongly coupled optical phonons in the ultrafast
		dynamics of the electronic energy and current relaxation in graphite}.
	\newblock \emph{\bibinfo{journal}{Phys. Rev. Lett.}}
	\textbf{\bibinfo{volume}{95}}, \bibinfo{pages}{187403}
	(\bibinfo{year}{2005}).
	
	\bibitem{HelmPRL2011}
	\bibinfo{author}{Winnerl, S.} \emph{et~al.}
	\newblock \bibinfo{title}{Carrier relaxation in epitaxial graphene photoexcited
		near the dirac point}.
	\newblock \emph{\bibinfo{journal}{Phys. Rev. Lett.}}
	\textbf{\bibinfo{volume}{107}}, \bibinfo{pages}{237401}
	(\bibinfo{year}{2011}).
	
	\bibitem{ElsaesserTrOpt2011}
	\bibinfo{author}{Breusing, M.} \emph{et~al.}
	\newblock \bibinfo{title}{Ultrafast nonequilibrium carrier dynamics in a single
		graphene layer}.
	\newblock \emph{\bibinfo{journal}{Phys. Rev. B}} \textbf{\bibinfo{volume}{83}},
	\bibinfo{pages}{153410} (\bibinfo{year}{2011}).
	
	\bibitem{Cavalleri_2013}
	\bibinfo{author}{Gierz, I.} \emph{et~al.}
	\newblock \bibinfo{title}{Snapshots of non-equilibrium {Dirac} carrier
		distributions in graphene}.
	\newblock \emph{\bibinfo{journal}{Nat. Mater.}} \textbf{\bibinfo{volume}{12}},
	\bibinfo{pages}{1119--1124} (\bibinfo{year}{2013}).
	
	\bibitem{Hofmann_2013}
	\bibinfo{author}{Johannsen, J.~C.} \emph{et~al.}
	\newblock \bibinfo{title}{Direct view of hot carrier dynamics in graphene}.
	\newblock \emph{\bibinfo{journal}{Phys. Rev. Lett.}}
	\textbf{\bibinfo{volume}{111}}, \bibinfo{pages}{027403}
	(\bibinfo{year}{2013}).
	
	\bibitem{Hofmann_2014Nano}
	\bibinfo{author}{Johannsen, J.~C.} \emph{et~al.}
	\newblock \bibinfo{title}{Tunable carrier multiplication and cooling in
		graphene}.
	\newblock \emph{\bibinfo{journal}{Nano Lett.}} \textbf{\bibinfo{volume}{15}},
	\bibinfo{pages}{326--331} (\bibinfo{year}{2015}).
	
	\bibitem{Cavalleri_2015}
	\bibinfo{author}{Gierz, I.} \emph{et~al.}
	\newblock \bibinfo{title}{Tracking primary thermalization events in graphene
		with photoemission at extreme time scales}.
	\newblock \emph{\bibinfo{journal}{Phys. Rev. Lett.}}
	\textbf{\bibinfo{volume}{115}}, \bibinfo{pages}{086803}
	(\bibinfo{year}{2015}).
	
	\bibitem{Cavalleri_2014}
	\bibinfo{author}{Gierz, I.}, \bibinfo{author}{Link, S.},
	\bibinfo{author}{Starke, U.} \& \bibinfo{author}{Cavalleri, A.}
	\newblock \bibinfo{title}{Non-equilibrium {Dirac} carrier dynamics in graphene
		investigated with time- and angle-resolved photoemission spectroscopy}.
	\newblock \emph{\bibinfo{journal}{Faraday Discuss.}}
	\textbf{\bibinfo{volume}{171}}, \bibinfo{pages}{311} (\bibinfo{year}{2014}).
	
	\bibitem{Hofmann_2014PRL}
	\bibinfo{author}{Ulstrup, S.} \emph{et~al.}
	\newblock \bibinfo{title}{Ultrafast dynamics of massive {Dirac} fermions in
		bilayer graphene}.
	\newblock \emph{\bibinfo{journal}{Phys. Rev. Lett.}}
	\textbf{\bibinfo{volume}{112}}, \bibinfo{pages}{257401}
	(\bibinfo{year}{2014}).
	
	\bibitem{CavalleriPRL2015Bilayer}
	\bibinfo{author}{Gierz, I.} \emph{et~al.}
	\newblock \bibinfo{title}{Phonon-pump extreme-ultraviolet-photoemission probe
		in graphene: anomalous heating of {Dirac} carriers by lattice deformation}.
	\newblock \emph{\bibinfo{journal}{Phys. Rev. Lett.}}
	\textbf{\bibinfo{volume}{114}}, \bibinfo{pages}{125503}
	(\bibinfo{year}{2015}).
	
	\bibitem{Gierz_2017PRB}
	\bibinfo{author}{Aeschlimann, S.} \emph{et~al.}
	\newblock \bibinfo{title}{Ultrafast momentum imaging of pseudospin-flip
		excitations in graphene}.
	\newblock \emph{\bibinfo{journal}{Phys. Rev. B}} \textbf{\bibinfo{volume}{96}},
	\bibinfo{pages}{020301(R)} (\bibinfo{year}{2017}).
	
	\bibitem{GierzPRB2017}
	\bibinfo{author}{Pomarico, E.} \emph{et~al.}
	\newblock \bibinfo{title}{Enhanced electron-phonon coupling in graphene with
		periodically distorted lattice}.
	\newblock \emph{\bibinfo{journal}{Phys. Rev. B}} \textbf{\bibinfo{volume}{95}},
	\bibinfo{pages}{024304} (\bibinfo{year}{2017}).
	
	\bibitem{MatsudaPRB17}
	\bibinfo{author}{Someya, T.} \emph{et~al.}
	\newblock \bibinfo{title}{Suppression of supercollision carrier cooling in high
		mobility graphene on {S}i{C}(0001)}.
	\newblock \emph{\bibinfo{journal}{Phys. Rev. B}} \textbf{\bibinfo{volume}{95}},
	\bibinfo{pages}{165303} (\bibinfo{year}{2017}).
	
	\bibitem{PhysRevB.101.035128}
	\bibinfo{author}{Caruso, F.}, \bibinfo{author}{Novko, D.} \&
	\bibinfo{author}{Draxl, C.}
	\newblock \bibinfo{title}{Photoemission signatures of nonequilibrium carrier
		dynamics from first principles}.
	\newblock \emph{\bibinfo{journal}{Phys. Rev. B}}
	\textbf{\bibinfo{volume}{101}}, \bibinfo{pages}{035128}
	(\bibinfo{year}{2020}).
	
	\bibitem{BauerPRLG2018}
	\bibinfo{author}{Rohde, G.} \emph{et~al.}
	\newblock \bibinfo{title}{{Ultrafast Formation of a Fermi-Dirac Distributed
			Electron Gas}}.
	\newblock \emph{\bibinfo{journal}{Phys. Rev. Lett.}}
	\textbf{\bibinfo{volume}{121}}, \bibinfo{pages}{256401}
	(\bibinfo{year}{2018}).
	
	\bibitem{Damascelli2019Sci}
	\bibinfo{author}{Na, M.~X.} \emph{et~al.}
	\newblock \bibinfo{title}{Direct determination of mode-projected
		electron-phonon coupling in the time domain}.
	\newblock \emph{\bibinfo{journal}{Science}} \textbf{\bibinfo{volume}{366}},
	\bibinfo{pages}{1231--1236} (\bibinfo{year}{2019}).
	
	\bibitem{GedikNatureComm}
	\bibinfo{author}{Sie, E.~J.}, \bibinfo{author}{Rohwer, T.},
	\bibinfo{author}{Lee, C.} \& \bibinfo{author}{Gedik, N.}
	\newblock \bibinfo{title}{Time-resolved { XUV ARPES} with tunable {24-33} ev
		laser pulses at 30 {meV} resolution}.
	\newblock \emph{\bibinfo{journal}{Nat. Commun.}} \textbf{\bibinfo{volume}{10}},
	\bibinfo{pages}{3535} (\bibinfo{year}{2019}).
	
	\bibitem{BauerNature2011}
	\bibinfo{author}{Rohwer, T.} \emph{et~al.}
	\newblock \bibinfo{title}{Collapse of long-range charge order tracked by
		time-resolved photoemission at high momenta}.
	\newblock \emph{\bibinfo{journal}{Nature}} \textbf{\bibinfo{volume}{471}},
	\bibinfo{pages}{490--493} (\bibinfo{year}{2011}).
	
	\bibitem{Chamon_2000}
	\bibinfo{author}{Chamon, C.}
	\newblock \bibinfo{title}{Solitons in carbon nanotubes}.
	\newblock \emph{\bibinfo{journal}{Phys. Rev. B}} \textbf{\bibinfo{volume}{62}},
	\bibinfo{pages}{2806--2812} (\bibinfo{year}{2000}).
	
	\bibitem{Cheianov_2009}
	\bibinfo{author}{Cheianov, V.~V.}, \bibinfo{author}{Fal'ko, V.~I.},
	\bibinfo{author}{Sylju\.asen, O.} \& \bibinfo{author}{Altshuler, B.~L.}
	\newblock \bibinfo{title}{Hidden {Kekul\'e} ordering of adatoms on graphene}.
	\newblock \emph{\bibinfo{journal}{Solid State Commun.}}
	\textbf{\bibinfo{volume}{149}}, \bibinfo{pages}{1499} (\bibinfo{year}{2009}).
	
	\bibitem{JohanssonPRB2010}
	\bibinfo{author}{Virojanadara, C.}, \bibinfo{author}{Watcharinyanon, S.},
	\bibinfo{author}{Zakharov, A.~A.} \& \bibinfo{author}{Johansoon, L.~I.}
	\newblock \bibinfo{title}{Epitaxial graphene on 6{H}-{S}i{C} and {L}i
		intercalation}.
	\newblock \emph{\bibinfo{journal}{Phys. Rev. B}} \textbf{\bibinfo{volume}{82}},
	\bibinfo{pages}{205402} (\bibinfo{year}{2010}).
	
	\bibitem{Takahashi2011}
	\bibinfo{author}{Sugawara, K.}, \bibinfo{author}{Kanetani, K.},
	\bibinfo{author}{Sato, T.} \& \bibinfo{author}{Takahashi, T.}
	\newblock \bibinfo{title}{Fabrication of {L}i-intercalated bilayer graphene}.
	\newblock \emph{\bibinfo{journal}{AIP Adv.}} \textbf{\bibinfo{volume}{1}},
	\bibinfo{pages}{022103} (\bibinfo{year}{2011}).
	
	\bibitem{ChanghuaPRL2021}
	\bibinfo{author}{Bao, C.} \emph{et~al.}
	\newblock \bibinfo{title}{{Experimental Evidence of Chiral Symmetry Breaking in
			Kekul\'e-Ordered Graphene}}.
	\newblock \emph{\bibinfo{journal}{Phys. Rev. Lett.}}
	\textbf{\bibinfo{volume}{126}}, \bibinfo{pages}{206804}
	(\bibinfo{year}{2021}).
	
	\bibitem{OttavianoPRB}
	\bibinfo{author}{Bisti, F.} \emph{et~al.}
	\newblock \bibinfo{title}{{Electronic and geometric structure of
			graphene/SiC(0001) decoupled by lithium intercalation}}.
	\newblock \emph{\bibinfo{journal}{Phys. Rev. B}} \textbf{\bibinfo{volume}{91}},
	\bibinfo{pages}{245411} (\bibinfo{year}{2015}).
	
	\bibitem{MauriNatPhys2012}
	\bibinfo{author}{Profeta, G.}, \bibinfo{author}{Calandra, M.} \&
	\bibinfo{author}{Mauri, F.}
	\newblock \bibinfo{title}{Phonon-mediated superconductivity in graphene by
		lithium deposition}.
	\newblock \emph{\bibinfo{journal}{Nat. Phys.}} \textbf{\bibinfo{volume}{8}},
	\bibinfo{pages}{131} (\bibinfo{year}{2012}).
	
	\bibitem{Devereaux_PRX2013}
	\bibinfo{author}{Sentef, M.} \emph{et~al.}
	\newblock \bibinfo{title}{Examining electron-boson coupling using time-resolved
		spectroscopy}.
	\newblock \emph{\bibinfo{journal}{Phys. Rev. X}} \textbf{\bibinfo{volume}{3}},
	\bibinfo{pages}{041033} (\bibinfo{year}{2013}).
	
	\bibitem{Tom_PRB2014}
	\bibinfo{author}{Kemper, A.~F.}, \bibinfo{author}{Sentef, M.~A.},
	\bibinfo{author}{Moritz, B.}, \bibinfo{author}{Freericks, J.~K.} \&
	\bibinfo{author}{Devereaux, T.~P.}
	\newblock \bibinfo{title}{Effect of dynamical spectral weight redistribution on
		effective interactions in time-resolved spectroscopy}.
	\newblock \emph{\bibinfo{journal}{Phys. Rev. B}} \textbf{\bibinfo{volume}{90}},
	\bibinfo{pages}{075126} (\bibinfo{year}{2014}).
	
\end{thebibliography}
\end{document}